\documentclass[
    ,final            
  ]
  {aipproc}

\layoutstyle{6x9}

\def\gta{\lower 0.5ex\hbox{$ \buildrel>\over\sim\ $}}
\def\lta{\lower 0.5ex\hbox{$ \buildrel<\over\sim\ $}}

\begin{document}

\title{Effect of Radiative Levitation on Calculations of Accretion Rates in White Dwarfs}


\author{P. Chayer}{
  address={Space Telescope Science Institute, 3700 San Martin Drive, Baltimore, MD, 21218, USA}
}

\author{J. Dupuis}{
  address={Canadian Space Agency, 6767 Route de l'A\'eroport, Saint-Hubert, QC, J3Y~8Y9, Canada}
}

\begin{abstract}

Elements heavier than hydrogen or helium that are present in the
atmospheres of white dwarfs with effective temperatures lower than
25,000 K, are believed to be the result of accretion. By measuring the
abundances of these elements and by assuming a steady-state accretion, we can derive the composition of the accreted matter and infer its source. The presence of radiative levitation, however, may affect the determination of the accretion rate. We present time-dependent diffusion calculations that take into account radiative levitation and accretion. The calculations are performed on C, N, O, Ne, Na, Mg, Al, Si, S, Ar, and Ca in hydrogen-rich white dwarf models with effective temperatures lower than 25,000 K and a gravity of $\log g = 8.0$. We show that in the presence of accretion, the abundance of an element supported by the radiative levitation is given by the equilibrium between the radiative and gravitational accelerations, unless the abundance predicted by the steady-state accretion is much greater than the abundance supported by the radiative acceleration.

\end{abstract}

\maketitle


\section{INTRODUCTION}

The accretion of circumstellar matter is one of the most plausible physical processes that can explain the appearance of metals observed in the atmospheres of several white dwarfs with effective temperatures lower than 25,000~K (see, e.g., Dupuis, Fontaine, \& Wesemael \cite{dfw93} and Koester \& Wilken \cite{koesterwilken06}). In this scenario and under the assumption of a steady-state, the matter that falls onto the surface of a white dwarf diffuses downward because of the gravitational settling, and at the same time, it is continuously replenished with new material in such a way that the observed abundances remain constant. This accretion-diffusion model can maintain appreciable traces of metals in the atmospheres of white dwarfs and account for the observed abundances. The discovery of circumstellar dusty and gas disks, or both, around the most metal-rich white dwarfs goes in the direction that these disks are the source of the accreted matter (see, e.g., Zuckerman \& Becklin \cite{zuckerman_becklin}; Reach et al. \cite{reach_etal05}; G\"ansicke et al. \cite{gansicke_etal06}; Farihi, Jura, \& Zuckerman \cite{farihi_etal09}; and Brinkworth et al. \cite{brinkworthetal09}). Spectroscopic optical observations of such white dwarfs suggest that the circumstellar disks are the remaining of extrasolar minor planets that have been tidally disrupted by the white dwarf (see, e.g., Zuckerman et~al. \cite{zuckerman_etal07}; Klein et~al. \cite{klein_etal10}; and Dufour et~al \cite{dufour_etal10}). 

The accretion-diffusion model shows that under the steady-state assumption, the abundances of metals observed in the atmospheres of cool white dwarfs can be related to the composition of circumstellar disks. For this reason, white dwarfs are excellent probe to determine the composition of debris disks, and ultimately, the composition of the extrasolar minor planets. The accretion-diffusion model is only valid, however, when no other physical mechanisms interfere with the accretion. In this paper, we demonstrate that the radiative levitation on elements such as carbon, aluminum and silicon in relatively cool DA models is important and cannot be neglected when one tries to determine the composition of the accreted matter. 

\section{Radiative Levitation}

\begin{figure}[th]
  \includegraphics[height=.40\textheight]{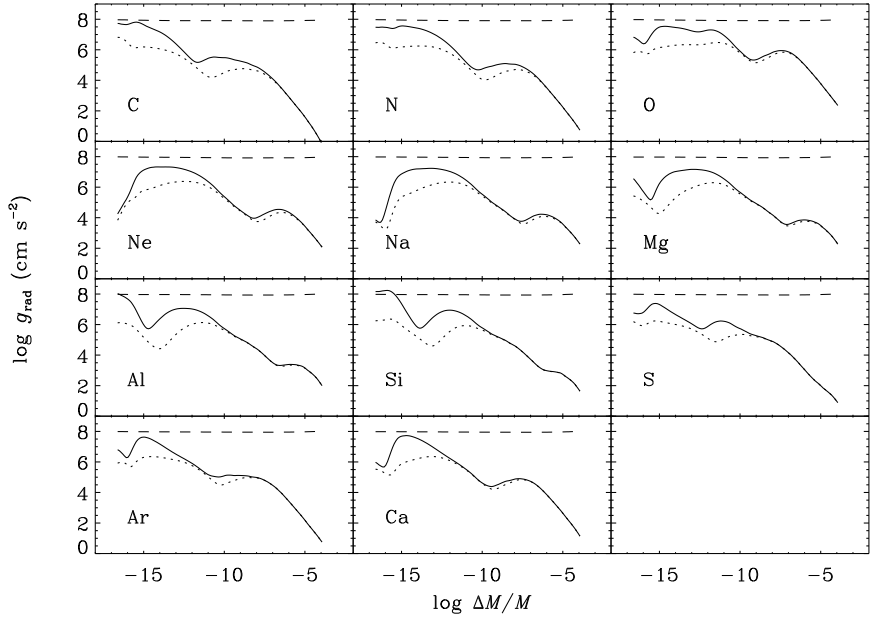}
  \caption{Radiative acceleration as a function of depth for 11 elements in a DA model with $T_{\rm{eff}} = 20$,000~K and $\log g = 8.0$. The solid curve gives the maximum radiative acceleration that can be transmitted to an element, i.e., in the regime where the lines are completely desaturated. The dotted curve gives the radiative acceleration for an abundance of mass fraction $X({\rm{A}}) = 10^{-2}$. The horizontal dashed curve gives the local effective gravity. \label{grad_profiles}}
\end{figure}

Figure~\ref{grad_profiles} shows the radiative acceleration as a function of depth for 11 elements in a DA model with $T_{\rm{eff}} = 20$,000~K and $\log g = 8.0$. These calculations are based on the work of Chayer et al. \cite{cfw95}.  The solid curve gives the maximum radiative acceleration that is transferred to an element when all its absorption lines are desaturated. On the other hand, by increasing the abundance the saturation of the lines increases and, as a result, the radiative acceleration decreases. This is illustrated by the dotted curve in Figure~\ref{grad_profiles}. The radiative acceleration in this case is computed for an abundance of mass fraction $X({\rm{A}}) = 10^{-2}$. The figure shows that aluminum and silicon have radiative accelerations that are greater than the gravitational acceleration (dashed curve). Consequently, these two elements can be maintained in the atmosphere of such a white dwarf by the radiative levitation.

\section{Accretion and Radiative Levitation}

In order to answer the question on how the radiative levitation affects the calculations of accretion rates, we have carried out two sets of 
time-dependent diffusion calculations. The first set includes diffusion in presence of accretion alone, and the second set includes diffusion in presence of both accretion and radiative levitation. The diffusion of an element in an white dwarf envelope in presence of accretion is followed by resolving the continuity equation with a boundary condition that acts as a source term at the surface of the star.  The radiative acceleration, on the other hand,  is considered as an additional term in the microscopic diffusion velocity that enters in the continuity equation. Examples of time-dependent diffusion calculations can be found in Chayer, Fontaine, \& Pelletier~\cite{chayer_etal97}, Dupuis et al.~\cite{dupuis_etal93}. and Vennes et al.~\cite{vennes_etal88}.

\begin{figure}[t]
  \includegraphics[height=.40\textheight]{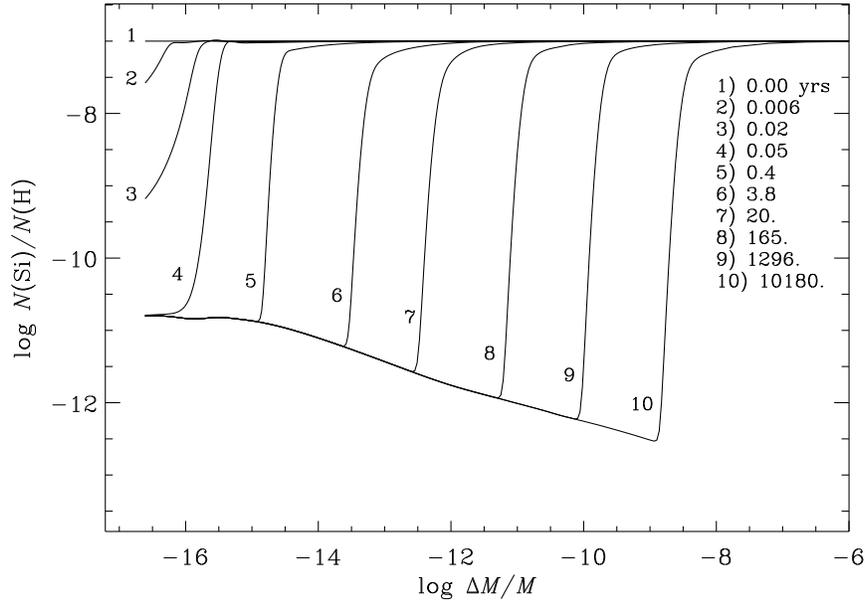}
  \caption{Evolving distribution of Si in a DA model in the presence of a weak accretion, but without considering the effect of the radiative acceleration. Silicon is accreted at a rate of $\dot{M}_{\rm{acc}} = 100$~g~s$^{-1}$ onto the surface of a DA model with $T_{\rm{eff}} = 20$,000~K and $\log g = 8.0$. Each profile is labeled by an increasing number that gives on the right hand side of the plot the corresponding time in years. We considered a uniform Si abundance of $\log N({\rm{Si}})/N({\rm{H}}) = -7.0$ as the initial condition. \label{accretion_nogr}}
\end{figure}

Figure~\ref{accretion_nogr} shows the diffusion of Si as a function of depth in presence of accretion in a DA model with $T_{\rm{eff}} = 20$,000~K and $\log g = 8.0$.  The calculations start with an homogeneous Si abundance corresponding to $\log N({\rm{Si}})/N({\rm{H}}) = -7.0$ (curve labeled by the number 1), and by considering a continuous accretion of Si onto the surface of the white dwarf at a rate of $\dot{M}_{\rm{acc}} = 100$~g~s$^{-1}$. The top of the envelope is at the Rosseland optical depth $\tau_{\rm{R}} = 2/3$. Although the calculations are performed for a period of $t = 10^6$ years, the figure shows the distribution of Si at 10 time steps from $t = 0$ to $t = 10$,180 years. The figure shows that under the influence of the gravitational settling, Si diffuses very rapidly from an abundance of $\log N({\rm{Si}})/N({\rm{H}}) = -7.0$ to an abundance corresponding to the steady-state value, i.e., the abundance for which there is an equilibrium between the accretion and the gravitational settling. The steady-state abundance at $\tau_{\rm{R}} = 2/3$ is  $\log N({\rm{Si}})/N({\rm{H}}) = -10.8$. At greater depths,  as Si continues to flow downward it reaches a steady state abundance at each depth. 
  
 \begin{figure}
  \includegraphics[height=.40\textheight]{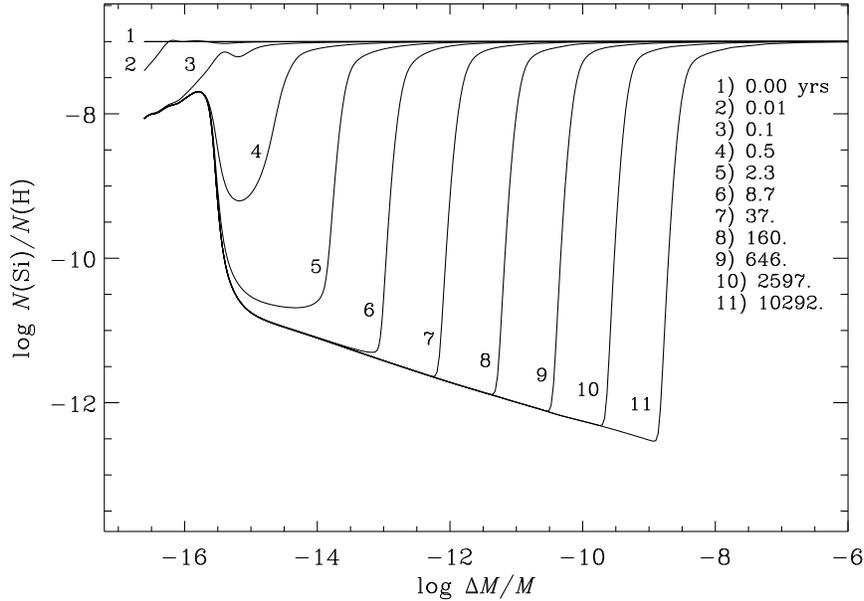}
  \caption{Same as Fig.~\ref{accretion_nogr}, but the radiative acceleration on silicon is taken into account. \label{accretion_gr}}
\end{figure}

Figure~\ref{accretion_gr} shows the evolution of the Si distribution as we have discussed in the previous paragraph, but this time we consider the effect of the radiative levitation. Silicon is still accreted at rate of $\dot{M}_{\rm{acc}} = 100$~g~s$^{-1}$. We have seen in Figure~\ref{grad_profiles} that Si is maintained in the upper layers of the atmosphere of a DA white dwarf with $T_{\rm{eff}} = 20$,000~K and $\log g = 8.0$. According to the radiative levitation theory, Si is supported because  the upward radiative acceleration is equal to the downward gravitational acceleration (Chayer et al.~\cite{cfw95}). By taking into account the radiative acceleration in the time-dependent diffusion calculations, we show in Figure~\ref{accretion_gr} that Si diffuses rapidly and then reaches the equilibrium abundance given by the radiative levitation.  A reservoir of Si builds up in the region where the radiative acceleration is greater than the gravitational acceleration. At deeper depths, however, where no radiative support is possible, Si flows downward and reaches a steady state that is almost identical to the one illustrated in Figure~\ref{accretion_nogr}. The slight difference is due to the radiative acceleration that slows down the diffusion of Si. In contrast with Figure~\ref{accretion_nogr}, the Si abundance on top of the envelope is no longer the abundance given by the steady state, but it is rather given by the radiative levitation. The Si abundance at $\tau_{\rm{R}} = 2/3$ is $N({\rm{Si}})/N({\rm{H}}) = -8.1$, which is about 500 times greater than the steady-state abundance when Si is accreted at a rate of $\dot{M}_{\rm{acc}} = 100$~g~s$^{-1}$. 

\begin{figure}
  \includegraphics[height=.40\textheight]{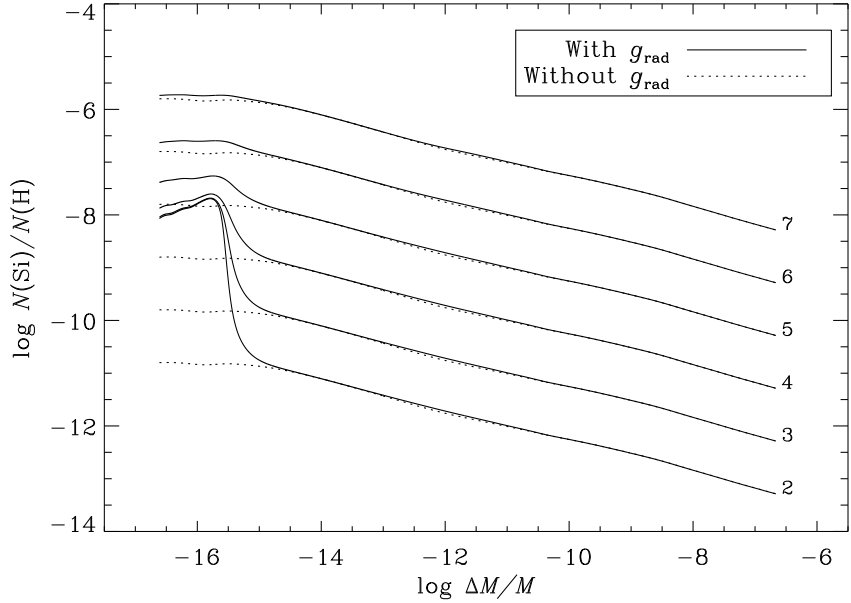}
  \caption{Comparison of the distributions of Si in a DA model in the presence of accretion computed at different accretion rates to models that include the radiative acceleration on Si (solid curve), and to models that do not include it (dotted curve). These distributions are obtained by following the diffusion of Si throughout the envelope of a DA model with $T_{\rm{eff}} = 20$,000~K and $\log g = 8.0$ for $10^6$ years. The curves are labeled by the logarithmic value of the accretion rate $\dot{M}_{\rm{acc}}$ in g~s$^{-1}$. \label{accretion_results}}
\end{figure}

We have computed the evolution of the Si distribution in a DA model when Si is accreted onto the surface of the star at a very low  accretion rate, $\dot{M}_{\rm{acc}} = 100$~g~s$^{-1}$. What happens to the Si distribution when Si is accreted onto the star at a higher rate? How the equilibrium abundance supported by the radiative levitation is affected by higher accretion rates? Figure~\ref{accretion_results} illustrates this situation by comparing the distributions of Si in our DA model when Si is accreted at different accretion rates. The distributions of Si are computed by taking the radiative acceleration into consideration (solid curve), and without taking it into consideration (dotted curve). The accretion rates range from $\dot{M}_{\rm{acc}} = 10^2$ to $10^7$~g~s$^{-1}$. Each curve in the figure is labeled by the logarithmic value of the accretion rate. The figure shows that the distributions of Si that are computed with and without the radiative levitation are almost identical for depths $\log \Delta M/M\ \gta {-15}$. However, in the regions where the radiative support becomes important and when Si is accreted with rates of $\dot{M}_{\rm{acc}}\ \lta 10^4$~g~s$^{-1}$, the abundance at the top of the envelope is not set by the steady-state abundance, but by the equilibrium abundance supported by the radiative levitation. On the other hand, when Si is accreted at higher rates, i.e., for accretion rates $\dot{M}_{\rm{acc}}\ \gta 10^5$~g~s$^{-1}$, the Si abundances at the surface of the star converge to the steady-state abundances. For these high accretion rates, the abundances are no longer given solely by the radiative levitation, but rather by the equilibrium between the accretion and the downward diffusion. 

\section{Conclusion}

Our findings demonstrate that the radiative levitation affects the observed abundances at the surface of DA white dwarfs when accretion is present. Although we have only showed time-dependent diffusion calculations on Si in one DA model ($T_{\rm{eff}} = 20$,000~K and $\log g = 8.0$), we have in fact carried out our investigation for a range of effective temperatures from 10,000~K to 25,000~K, one gravity $\log g = 8.0$, and 11 elements. With this in mind, we can report that among the 11 elements that we have analyzed, C, Al and Si are the elements that are susceptible to the radiative support. If these elements are observed in the atmospheres of DA white dwarfs with 17,000~K $\lta T_{\rm{eff}}\ \lta 25$,000~K, their abundances could be interpreted according to three scenarios: 1) when there is no accretion, the observed abundances should correspond to the radiative equilibrium abundances; 2) when there is accretion with low accretion rates, the observed abundances should still be the radiative equilibrium abundances; 3) when there is accretion with high accretion rates, the observed abundances should approach the steady-state abundances. The non-detection of these elements, however, could indicate that they have left the atmosphere of the white dwarf during the earlier phases of its evolution. 

For a direct application of the current model to a handful of DA white dwarfs with $T_{\rm{eff}}\ \lta 25$,000~K, see Dupuis, Chayer, \& H\'enault-Brunet (these proceedings).


\begin{theacknowledgments}
P.C. is supported by the Canadian Space Agency under a Public Works Government Services Canada contract.
\end{theacknowledgments}



\bibliographystyle{aipproc}   

\end{document}